\documentclass[%
 reprint,
 amsmath,amssymb,
 aps,
]{revtex4-2}

\usepackage[utf8]{inputenc}    
\usepackage[T1]{fontenc}
\usepackage{textcomp}          
\usepackage{newunicodechar}    
\newunicodechar{ }{~} 
\usepackage{verbatim}
\usepackage{geometry}
\usepackage{amsmath,amsfonts,amssymb,color,dsfont}
\usepackage{graphicx}
\usepackage{multirow}

\begin{document}

\title{
Rescuing 331 bileptons from the Landau pole}

\author{Stefano Morisi}
% \altaffiliation[Also at ]{Physics Department, XYZ University.}%Lines break automatically or 
\email{stefano.morisi@unina.it}
\author{Giulia Ricciardi}%
 \email{giulia.ricciardi2@unina.it}
\affiliation{%
 \mbox{Dipartimento di Fisica ``Ettore Pancini''\text{, }Università degli Studi di Napoli ``Federico II''%\text{, }Complesso Univ. Monte S. Angelo\text{, } I
}\\
 \mbox{INFN - Sezione di Napoli, Complesso Univ. Monte S. Angelo, I-80126 Napoli, Italy}\\
}%

\author{Giovanna Paola Perdonà}
% \altaffiliation[Also at ]{Physics Department, XYZ University.}%Lines break automatically or 
\email{giovannapaola.perdona@uniroma1.it}
\affiliation{%
 \mbox{Dipartimento di Fisica, Sapienza Università di Roma}\\
 \mbox{INFN - {Sezione di Roma, Piazzale Aldo Moro, 2 - 00185 Roma RM, Italy}}
}

%\date{\today}

\begin{abstract}
\noindent
Among the particles being searched for at the LHC  beyond the Standard Model are bileptons, which are doubly charged gauge bosons.
\noindent
Bileptons are predicted by several Standard Model extensions, including the so-called 331 models with $\beta = \sqrt 3$. The minimal formulation of these  models is  generally plagued by a low-energy Landau pole, which can undermine current predictions for TeV-scale bileptons. 
We analyze this issue and investigate the possibility to  shift the Landau pole at higher energy scales by extending the troublesome minimal 331 models.
  \end{abstract}

\maketitle
\section{Introduction}

There are several and well known reasons for extending the Standard Model (SM) $SU(3)_c\times SU(2)_L\times U_Y(1)$ group. An interesting possibility 
is to enlarge the gauge group $SU_L(2)$ to $SU_L(3)$ or more precisely $SU(2)_L\times U_Y(1)$
to $SU(3)_L\times U_X(1)$ where $X$ is a generalized hypercharge. 
In\,\cite{Singer:1980sw} it was suggested for the first time to embed the three families of fermions  into irreducible representations of the gauge group $SU_L(3)\times U_X(1)$. However, the first complete model based on $SU_L(3)\times U_X(1)$, and  hereafter named 331 (the first 3 refers to $SU_c(3)$)
has been proposed by\,\cite{Pisano:1992bxx} and \cite{Frampton:1992wt}. Quite a  large number of works has followed  and giving a complete list of all of them is beyond the scope of the present paper\,\footnote{For a complete introduction to 331 models and their features we refer e.g. to \cite{Buras:2012dp}.}. 

\noindent
From the phenomenological point of view, 331 models predict the existence of new fermions and gauge bosons beyond the Standard Model (BSM). The  gauge properties of the specific 331 model depend on an arbitrary parameter, denoted as $\beta$.
\noindent
In the case $\beta=\sqrt{3}$, one extra gauge boson is neutral and  two  have charges $Q=\pm 1$; we call them $Z', {W'}^\pm$, respectively. The remaining two extra gauge bosons are doubly charged and are denoted as $Y^{\pm\pm}$. Since they couple with two standard leptons they are called bileptons.
\noindent
From the  perspective of collider physics, the presence of such a bilepton is one of the most interesting features of these 331 models.
The existence of bileptons has been actively investigated at LHC, bringing about lower bounds on their masses~\cite{Nepomuceno:2019eaz,Calabrese:2023ryr}. Several phenomenological studies have explored their possible signatures at LHC (e.g.~\cite{Corcella:2017dns,Corcella:2018eib}).
\noindent
The downside of 331 models with $\beta=\sqrt{3}$ is the well-known presence of a Landau pole at low-energy scales.  Already in the 1950s Landau realized  that the QED running coupling diverges—developing a pole—at a finite momentum scale. 
 The presence of such a divergence, known as the Landau pole, is not limited to QED but poses a potential problem for a generic quantum field theory in the perturbative regime.
 The modern perspective on this issue involves several arguments. 
 One  key point is that the Landau pole arises within perturbation theory, which cannot be expected to reliably represent behaviors related to large or even infinite couplings. Real insight can only be gained using non-perturbative approaches. It is not unreasonable to assume that physical observables of a quantum field theory are well-behaved even when the coupling diverges at the Landau pole and beyond. This possibility has been explored and several models  have been identified where this behavior is realized (see e.g. \cite{Romatschke:2022llf} and references therein). 
 Another perspective emerges when we examine the energy scales of the Landau pole. 
In the Standard Model the Landau pole is above the Planck scale, thus it becomes of no practical significance to study the SM in that regime. This leads  to view theories with a Landau pole as UV-incomplete or effective theories. The presence of a Landau pole is taken as an evidence of  an energy scale at which the theory breaks down and
new physics must intervene.
 This interpretation becomes obviously more concerning when the scale of the Landau pole approaches energies accessible to present or near future experiments. 
 Since the  running of the couplings changes according to the matter content of the models, it is necessary to investigate SM extensions for the presence of a Landau pole.
\noindent
In the case of 331 models, the scale of the Landau pole depends on the value of the parameter $\beta$.
Minimal 331 extensions with $\beta = 1/\sqrt{3}$ exhibit a Landau pole in the $U(1)$ coupling at very high energies (much like the Standard Model), 
while setting $\beta = \sqrt{3}$ 
lowers the Landau pole to about $ 4$ TeV. 
Such a  low energy divergence can be troublesome for bileptons detection in a near future by LHC. 
It is interesting both from the phenomenological and theoretical point of view to investigate
how to build extensions of the minimal 331 model with $\beta = \sqrt{3}$ where the Landau pole occurs at  higher energy scales \cite{Dias:2004dc,Martinez:2006gb,Doff:2023bgy,Barela:2023oyp}.\\
\noindent
In this paper we address the Landau pole problem  considering next-to-minimal 331 extensions.

\section{General features of 331 models}
\label{subsec:matter}
In 331 models we extend the  Standard Model gauge group to $SU(3)_C\times SU(3)_L\times U(1)_X$.
The number of the new gauge bosons is fixed by the
number of $SU(3)_L$ generators, namely 8 (compared to the 3 of $SU(2)_L$ in the SM). Since  there are  3 diagonal generators,  the hypercharge $Y$ depends on an arbitrary parameter, denoted as $\beta$,  which defines the gauge properties of the specific 331 model.
The electric charge is defined as 
\begin{equation}
    \hat{\mathcal{Q}}=  \hat T^3 + \beta \hat T^8+X{\mathds{I}}
\label{hyper:ID}
\end{equation}
where  $X$ is the quantum number associated with $U(1)_X$ and ${\mathds{I}}$ is the identity matrix.
This expression generalizes the Gell-Mann Nishijima relation, where the $X$-charge is related to the SM hypercharge $Y$. We have
\begin{equation}
\frac{Y}{2} = \beta T^8 + X{\mathds{I}}.
\end{equation}
We classify 331 extensions  according to the value chosen for  $\beta$. \\
\noindent
One of the five new gauge bosons is neutral, independently of the value of $\beta$, and it is generally denoted as $Z'$. 
 The charges of the remaining  extra four gauge bosons depend on $\beta$.\\
 \noindent
Other than extra gauge bosons, 331 extensions predict extra fermions, 
whose charges are also given by  Eq.  \eqref{hyper:ID} and depend on $\beta$.
Unlike the gauge sector, their number depends on the matter assignment, which is model dependent. The smallest irreducible representation of $SU_L(3)$ is the  triplet. In analogy with the Standard Model, we associate any left-handed
quark and lepton family  to a $SU_L(3)$ triplet, or anti-triplet, representation. A crucial requirement for 331 extensions is gauge anomaly cancellation. It  constrains not only the matter content of the theory, but  also sets some criteria on how to assign quark and lepton families into
triplets and/or anti-triplets of the $SU_L(3)$ group.\\
\noindent
Let us name $N_q$ and $N_\ell$  the number of quark and lepton families, respectively.
Indicating with  $N_q^{3(\bar{3})}$ and $N_\ell^{3(\bar{3})}$  the number of quark and lepton families  in the $\boldsymbol{3}$  ($\boldsymbol{\bar{3}}$)  representations of $SU_L(3)$, 
 we can write 
\begin{align}
N_q &= N_q^3 + N_q^{\bar{3}} & \nonumber \\
N_\ell &= N_\ell^3 + N_\ell^{\bar{3}}.
\label{eqal:sum}
\end{align}
\noindent
Some general considerations are in order.
\begin{itemize}
\item[-]
The charges of the first two components of triplet (or antitriplet) must match the charges of
the corresponding SM fields. This matching fixes uniquely the $X$-values for the SM fermions as
functions of $\beta$. 
The non-standard fields  have the same $X$-values of the other members of
the respective (anti)triplet, because of gauge invariance.
Hence each quark (or lepton) family in a $\boldsymbol{3}$ representation    shares the same $X$-value and the same happens for each quark (or lepton) family in a $\boldsymbol{\bar{3}}$ representation. 
This constraint does not need to apply when we add triplets (or antitriplets) composed solely of left-handed BSM fermions to the matter content. However, by assumption, we choose to retain this constraint. 
In analogy to Standard Model, each non-standard left-handed quark needs a right-handed  component  to maintain $SU(3)_C$ vector-like couplings. Obviously, that does not apply to leptons. However, for the anomaly cancellation and in analogy with the SM, we assume that  each left-handed non-standard charged lepton has a right-handed counterpart of the same electric charge. 
 Neutral leptons do not participate to anomaly cancellation, therefore we do not need to assume  a right-handed counterpart.

Summarizing, on the matter content we make two assumptions: i)  
each $SU(3)_L$ charged fermion has a right-handed  counterpart ii) all exotic families are replicas of the standard ones.\\
Under these assumptions, 
the requirement of anomaly cancellation  implies  the relations: \cite{Calabrese:2023ryr} 
\begin{align}
N_F &\equiv  N_q = N_\ell, \label{e:familynumber1}\\
N_q^3 &= \frac{2N_F - N_\ell^3}{3},  
\label{e:familynumber}
\end{align}
where $N_{q(\ell)}$ is the number of quark (lepton) families.
\item[-]
Let us observe that the colour sector of the theory, namely $SU_c(3)$, 
does not change with respect to the SM. BSM quarks (also indicated as exotic) have colour as the SM ones. A necessary (but not sufficient) condition for QCD to be  
asymptotically free, which constrains 331 models' matter content, is 
 \begin{equation}     
n_{\text{quark}}<\frac{33}{2}. \label{eq:color} \end{equation} 
In the SM  there are three quark families, each family has two components, then  $n_{\text{quark}}=6 $, which respects the limit \eqref{eq:color}.
In $SU_L(3)$,
since we are assuming that each left-handed quark family is in a triplet (or antitriplet),  we have
\begin{equation}
 n_{\text{quark}}= 3 N_F \,.  
\end{equation}
Therefore from relation (\ref{eq:color}) it follows 
the limit $N_F \leq 5$. We note that, contrary to what is frequently reported in literature, $N_F$ is not necessarily a multiple of the colour number and can be $N_F=3,4,5$ \footnote{Let us notice that the $N_F=5$ value is excluded by the $SU(3)_c$ conformal window while the case $N_F = 4$ is still under discussion (see e.g. \cite{Aoki:2012eq,Mickley:2025mjj}).}. 
\item[-]
An additional, quite common, assumption is that the exotic gauge bosons of the 331 model have  integer charge values, in analogy to the SM ones, which is obtained by setting  \begin{equation}
    |\beta| = n/\sqrt{3}
\end{equation} where $n$ is an odd integer number. 
Another constraint for 331 models is obtained  by singling out the coupling with $Z$ boson and by matching with the corresponding
Standard Model coupling. This results in a condition on the gauge couplings of the two theories, which remain real 
if 
$ |\beta | <  1.8 $ holds (see \cite{Calabrese:2023ryr} for details).
Therefore, models with $n>3$ are excluded and $\beta$ is constrained to the values  $\beta=\pm1/\sqrt{3},\pm \sqrt{3}$.
\end{itemize}
\noindent
In the case $|\beta|=1/\sqrt{3}$ the BSM quarks have SM charges $-1/3$ and $2/3$; there are also 3 BSM neutral bosons (including $Z'$) and  charged BSM boson $W^{'\pm}$.
In the case $|\beta|=\sqrt{3}$ the extra quarks have non SM charges (4/3 and 5/3). The  charged  BSM bosons $W^{'\pm}$ are accompanied by  double charged bosons $Y^{\pm\pm}$, that can decay into a pair of same charge SM leptons, giving an interesting signature for LHC \cite{Pisano:1992bxx,Frampton:1992wt,Corcella:2017dns,Calabrese:2023ryr}.
They are indicated as  bileptons,
because they do not couple at tree level 
to SM quark pairs.
Therefore the case $|\beta|=\sqrt{3}$ presents more exotic features compared to the case $|\beta|=1/\sqrt{3}$
and seems more distinctive for LHC detection. In the following sections we will study extensions of the so-called minimal 331 model with $\beta = \sqrt{3}$ .

\section{Minimal 331-model %($m331$) 
}
\label{minimal331}
\noindent
We consider  331 models where  $\beta=\sqrt{3}$. All 331 models are constrained by relations \eqref{e:familynumber1} and  \eqref{e:familynumber}. 
In the case $N_F=3$, one possible choice 
of representations  is 
\begin{equation}
N^3_q = 2, \quad N^{\bar{3}}_q = 1, \quad N^3_\ell =  0, \quad N^{\bar{3}}_\ell = 3.
 \label{e:memchoice}
\end{equation}
The $SU_L(3)$ triplet (or anti-triplet) has SM fermions occupying the first two components and a BSM fermion the third one. Since according to \eqref{e:memchoice}
 the lepton multiplets transform as $\bar{3}$ under $SU_L(3)$, they  take the general form 
$(\ell_L,-\nu_L^\ell,\xi^\ell_L)$, where the index $\ell$ runs on the 3 families and  $\xi^\ell_L$ is a left-handed BSM lepton. We note that  each $\xi^\ell_L$ has the same electric charge of $\ell_R^c$ ($c$ stands for conjugate), namely $Q(\xi^\ell_L)=+1$, as well as the same $X$-value, and it belongs to the same 331 representation. The economical identification $\xi^\ell_L\equiv 
\ell_R^c$   is the assumption which characterizes the so-called {\it minimal 331 model}. 
\noindent
Let us observe that  we cannot  repeat the assumption  in the quark sector, since all BSM quark fields have unconventional electric charges ($-4/3$ or $5/3$). The matter content of the minimal 331 model is summarized in Table\,\ref{tab:minimal331}.
The gauge symmetry of  331 models is spontaneously broken in two steps, to the Standard Model and the electroweak symmetries, namely
\begin{align}\label{331breakingII}
&SU_L(3)\times U_X(1) \xrightarrow[]{\mu_{331}} SU_L(2)\times U_Y(1) \nonumber \\
&SU_L(2)\times U_Y(1) \xrightarrow[]{\mu_{\rm EW}}  U_{\rm e.m.} 
\end{align}
at  the 331   and electroweak scales, $ \mu_{331}$ and  $\mu_{\rm EW}$, respectively. Hence the matter content 
 is completed by three scalar $SU_L(3)$ triplets $\chi,\rho, \eta$. They represent the minimal set of Higgs bosons necessary to break the symmetry for the different
values of $\beta$ and $X$ charges in 331 models \cite{Diaz:2003dk}. Their $X$ values in the 331 minimal model are reported in Table \ref{tab:minimal331}.\\
\noindent
The "heavy" scalar triplet $\chi$ acquires a vacuum expectation value (VEV) $\langle \chi \rangle$, namely
\begin{equation}
\langle \chi \rangle = \begin{pmatrix} 0 \\ 0 \\ \frac{u}{\sqrt{2}} \end{pmatrix}.
\end{equation}
It spontaneously breaks the 331 symmetry group down to the SM one at the scale $\mu_{331} \sim u$
\begin{equation}\label{331breaking}
SU_L(3)\times U_X(1) \xrightarrow[]{\mu_{331}} SU_L(2)\times U_Y(1),  
\end{equation}
giving masses to the five BSM gauge bosons. Among them, the bileptons \( Y^{\pm\pm} \), doubly charged gauge bosons, are a distinctive feature of this model. They are called \textit{bileptons} because they carry lepton number \( L = \pm 2 \) and couple only to standard lepton pairs, not to standard quarks.
  
\noindent
The mass of the bilepton $Y$ depends  on both  $u$ and the $SU(2)_L$ coupling $g_{2L}$, since \cite{Buras:2012dp}
\begin{equation}  
m_Y \simeq u\,\frac{g_{2L}}{2} .
\label{e:bileptmass}
\end{equation}
This relation is derived under the assumption that the VEV scale $u$ is much larger than the VEVs  responsible of the SM symmetry breaking (see below).

A lower bound $m_Y \sim 1.3$ TeV at $90\%$C.L for the vector bilepton mass has been derived in \cite{Calabrese:2023ryr}, recasting the   ATLAS search  for the doubly charged Higgs bosons in \cite{ATLAS:2022pbd}. In  the analysys \cite{Calabrese:2023ryr}  the multi-lepton cross section of scalar bilepton has been rescaled to the vector one. A similar bound  has also been obtained through the use of renormalization group equations \cite{Coriano:2020iiz}.

%A lower bound  of  $\sim 1.3$ TeV at $90\%$C.L. \cite{Calabrese:2023ryr} can be inferred by   experimental constraints (see e.g. \cite{ATLAS:2022pbd} and \textcolor{red}{through the use of renormalization group equations \cite{Coriano:2020iiz}}). 

By using the value \cite{ParticleDataGroup:2024cfk}  $g_{2L}(M_Z)=\sqrt{4 \pi \alpha_{2L}(M_Z)}\simeq 0.65$ and assuming the lower bound   $m_Y\gtrsim  1.3$\,TeV
one finds 
\begin{equation}\label{limit331} 
    \mu_{331}\sim u \, \gtrsim \, 3850\,{\rm GeV}.
\end{equation}
All extra quark fields $D^{1,2},T$ acquire mass through $\langle \chi \rangle$, therefore their masses  are defined by $u$ and by the Yukawa couplings. The latter are free parameters with upper bounds dictated by perturbativity and vacuum stability. It is natural to expect the values of these masses to lie around the $u$ scale or below, with an approximate lower bound at the electroweak scale.   
\noindent
The $SU_L(2)\times U_Y(1)$ is spontaneously broken to $U(1)_{e.m.}$ by the VEVs of the additional two scalar  triplets:
\begin{equation}
\langle \rho \rangle = \begin{pmatrix} 0 \\ \frac{v}{\sqrt{2}} \\ 0 \end{pmatrix},\qquad
\langle \eta \rangle = \begin{pmatrix} \frac{v'}{\sqrt{2}}\\0 \\ 0 \end{pmatrix},
\end{equation}
where the VEVs are related to the SM one $v_{EW}$ by $v^2+v'^2=v_{EW}^2= (246\, \mathrm{GeV})^2\equiv \mu_{EW}^2$. 
 Since the EW symmetry breaking should occur at a lower scale, we expect the hierarchy $u\gg v,v',v_{EW}$. 
 We summarize the scalar content of the minimal 331 model in table\,\ref{tab:scalar}.
\begin{table}[h!]
    \centering
\begin{tabular}{|c|c|c|}
        \hline
        Scalar & $SU_L(3)$ & $U_X(1)$ \\
        \hline 
        $\chi$ & 3 &  $+1$\\
        $\rho$ & 3 & 0\\
        $\eta$ & 3 & $-1$\\
        \hline
\end{tabular}
    \caption{Scalar content of minimal 331 model.}
    \label{tab:scalar}
\end{table}
At this passage, three other generators are broken, which leads us to the three
massive SM gauge bosons and one massless gauge boson
(the photon). \\
\noindent
The scalar sector includes singly and double charged states as well as pseudoscalar and scalar neutral states \cite{Liu:1993gy, Tonasse:1996cx, Tully:1998wa, Nguyen:1998ui, Foot:1992rh, Buras:2012dp, Pinheiro:2022bcs}. 
The arbitrariness of the Higgs potential constants adds further complications to its assessment.

\subsection{The Landau Pole in the minimal 331 model}
\label{section31}

The gauge coupling $g_i$ associated to the gauge groups $U(1),\,SU(2),\,SU(3)$,  %in\,(\ref{331breakingII}),
or equivalentely 
the couplings $\alpha_i \equiv g_i^2/4\pi$, are not constant and their values depend on the energy scale $\mu$.
In perturbation theory, the  expression  for the running couplings $\alpha_i(\mu)$ given by the renormalization group in the so-called leading logarithm approximation is
\begin{equation}
\alpha_i(\mu) = \frac{\alpha_i(\mu_0)}{ 1 -\frac{b_i}{2\pi}\alpha_i(\mu_0)\log\left( \frac{\mu}{\mu_0} \right)} + \mathcal{O}(\alpha_i^2)\,, \label{e:alpharunning}
\end{equation}
where $\mu_0$  is some fixed reference value and $b_i$ the first coefficient of the perturbative expansion of the $\beta$ function. 
When  $b_i>0$,
  the coupling diverges at some energy scale (Landau pole).

\noindent
For a 
gauge theory with a semisimple gauge group and multiplets of fermions and scalars it is well know that 
\cite{Diaz:2005bw,Jones:1981we}:
\begin{align}
b_i &= \frac{2}{3}\sum_{fermions}\text{Tr}(T_aT_a) + \nonumber \\
  &+ \frac{1}{3}\sum_{scalars}\text{Tr}(T_aT_a) - \frac{11}{3}C_2, \label{e:coefficients}
\end{align}
where $C_2$ and $T_a$ are respectively the Casimir and the generators of the gauge group. The value of $b_i$, and therefore the scale of the Landau pole, critically depend on the matter content of the model. More precisely, the dependence is on how many fermions and scalars are active at the energy under scrutiny.
  Since $C_2=0$ for the $U(1)$ gauge group, 
 the  $b$ coefficient is positive, hence the $U(1)$ coupling   always exhibits a Landau pole.
\noindent
In the case of the SM gauge group  $ SU_L(2)\times U_Y(1)\times SU_c(3)$ we have:
\begin{eqnarray}
    b_Y &=& \frac{20}{9} N_F + 
    \frac{1}{6} N_H \nonumber \\ 
    b_{2L} &=& -\frac{22}{3}+\frac{4}{3} N_F + 
    \frac{1}{6} N_H \nonumber \\
     b_c  &=& -11 +\frac{4}{3} N_F, 
     \label{formuleg}
\end{eqnarray}
where $N_F$ is the number of families, $N_H$ is the number of Higgs doublets and the coefficients $b_Y,\,b_{2L},\,b_c$ correspond  to the coupling constants
$\alpha_Y,\,\alpha_{2L},\,\alpha_c $ of  $SU_L(2)$, $U_Y(1)$ and $SU_c(3)$, respectively.
We can
compare the $b_i$ values in the SM and in the so-called Two Higgs Doublet Model (2HDM), that differs from the SM for the presence of an additional Higgs doublet. We have
 \hspace*{-2cm}
\begin{eqnarray}
\label{bSM}
&&b_Y=\frac{41}{6}  \quad
     b_{2L}=-\frac{19}{6}  
     \quad
 b_c=-7\quad {\rm SM}
\label{formrsm}\\
\label{bSM2}
 &&b_Y=7 \quad
   b_{2L}=-3   \quad  b_c=-7 \quad{\rm 2HDM}
\label{formrsm1}
\end{eqnarray}
We underline that we are not employing the  fairly common $SU(5)$ normalization which sets $g_1^2
\equiv (5/3)g_Y^2$ and 
 would lead, for instance, to $b_1=b_Y \,3/5=41/10$ in the SM case. 

\noindent
In all 331 models the two scales $\mu_{331}$ and  $\mu_{\mathrm EW}$ delimitate energy regimes with different  symmetries and matter content, both of which affect the running.\\
\noindent
 The SM gauge group  dictates the running behavior of 
 the gauge couplings $g_Y,\,g_{2L}$ from the electroweak scale $\mu_{\rm EW} \sim M_Z$ up to the $\mu_{331}$ scale.
Above $\mu_{311}$ the heavy degrees of freedom become on-shell. The different matter content  and the  unbroken 331 gauge symmetry  lead the running of $g_X,\,g_{3L}$
above the  $\mu_{331}$ scale. The matching condition at $\mu_{331}$ is
given by\,\cite{Buras:2012dp}
\begin{align}
&g_{2L}(\mu_{331}) = g_{3L}(\mu_{331}) \,,\nonumber \\
 &\frac{1}{g_X^2(\mu_{331})} =   %\left.  
 \frac{1}{6}\left(\frac{1}{g_Y^2(\mu_{331})} - \beta^2\frac{1}{g_L^2(\mu_{331})}\right).
 %\right|_{\mu=\mu_{331}}.
    \label{eq:matching1}
\end{align}
It yields the value of $g_X^2 $ at the scale $\mu_{331}$ as a function of $g_Y^2$ and $g_L^2$ at the same scale. The latter values are obtained 
by running $\alpha_Y$ and $\alpha_L$ 
to $\mu_{0} = \mu_{331}$ from $\mu_{EW} \sim M_Z$, 
given the measured values \cite{ParticleDataGroup:2024cfk}
\begin{align}\label{initial}
    \alpha_Y^{-1}(M_Z)&\simeq 98.341\,,\nonumber \\
    \alpha_{2L}^{-1}(M_Z)&\simeq 29.589\,,\\
    \alpha_c^{-1}(M_Z)&\simeq 8.42\,. \nonumber
\end{align}
\noindent
The matching condition (\ref{eq:matching1}) gives the dependence of $\alpha_X^{-1}(\mu_{331}) $ on  $\mu_{331}$, which is clearly governed  by the running of $\alpha_Y$ and $\alpha_L$ from $\mu_{EW} \sim M_Z$
to $\mu_{331}$.  
The scale $\mu_{331}$ satisfies both the experimental lower bound 
\eqref{limit331} and the condition $\alpha_X^{-1}(\mu_{331})\ge 0$, but it is otherwise free.

\begin{figure}[tbh!]
\centering
 \includegraphics[width=.9\linewidth]{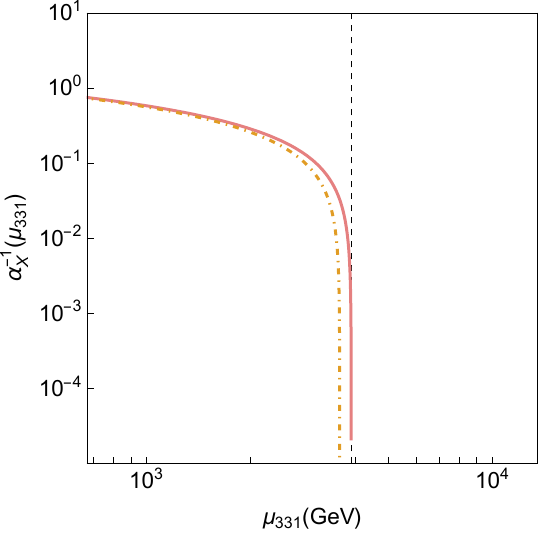}
  \caption{ We report the dependence of $\alpha_X^{-1}$ on the energy scale $\mu_{331}$.  The dot-dashed and continuous lines correspond respectively to the cases where one and two Higgs doublets are on-shell below $\mu_{331}$. The vertical dashed line represents the Landau pole at $3908$ GeV. 
  }
\label{figpole1}
\end{figure}
 \noindent
 To determine the running behavior of $\alpha_X^{-1}$, and hence its dependency on the matching scale $\mu_{331}$,  we need to clearly define  the matter content both above and below $\mu_{331}$.
In the scalar sector we expect that at least one of the three Higgs doublets  has  mass $\simeq \mu_{331}$ and contributes to the running only above the $ \mu_{331}$ scale.
We assume that
the heavy Higgs, the exotic gauge fields $W',Z',Y$ and the exotic fermions $D^1,\,D^2,\,T$  have masses $\simeq \mu_{331}$, hence they are not active below that scale.
 One of the other Higgs doublets can be  identified with the SM Higgs, leaving  two alternatives for the remaining Higgs doublet:
\begin{itemize} 
 \item[i)]
 "heavy", with mass around $\mu_{331}$. Only the SM Higgs   contributes to the $b_i$-coefficients below $\mu_{331}$.  \item[ii)]  
 "light", with mass between  the electroweak scale and $\mu_{331}$.
 The second Higgs doublet  affects the running of the couplings below $\mu_{331}$, mimicking the running of  the 2HDM in that region.
\end{itemize}
In  figure\,(\ref{figpole1}) we show 
$\alpha_X^{-1}(\mu_{331}) $
in case i), dot-dashed line, 
 and 
ii), continuous line.
A Landau pole arises when $\alpha_X^{-1}(\mu_{331})$
becomes zero, which occurs at 
at $\mu_{331}=3620$ GeV (case i)) 
and at 
$\mu_{331}=3908$ GeV (case ii)). These value set the upper limit 
of the 331 breaking scale $\mu^{\rm max}_{331}$. Thus we have:
\begin{itemize} 
 \item[i)]
$\mu_{331} \lesssim \mu^{\rm max}_{331}=3620  $ GeV
\item[ii)]$\mu_{331}\lesssim \mu^{\rm max}_{331}=3908$ GeV
 \end{itemize}
\noindent
Given the lower bound (\ref{limit331}), namely
$\mu_{331}\gtrsim 3850$ \, GeV,
 case i) is excluded by experimental data.\\ \noindent The only possibility remains case ii), which assumes two Higgs of mass  lower than $\mu_{331}$. 
The scale $\mu_{331}$ is constrained in the narrow range
\begin{equation}
 3850\,{\rm GeV}\lesssim \mu_{331}\lesssim 3908\,{\rm GeV}\,. \label{e:range}  
\end{equation}
\noindent
According to eqs.  \eqref{e:bileptmass} and \eqref{limit331}, the bilepton mass cannot exceed $\mu^{\rm max}_{331}$. \\
\noindent
At the upper limit $\mu^{\rm max}_{331}$, $\alpha_X^{-1}$ hits  the Landau pole.
Let us observe that, by definition of the mixing angle $\theta_W $, we can write
\begin{equation}
 \alpha_Y^{-1}(M_Z) =(1-\sin^2\theta_W) \alpha_{\rm em}^{-1}(M_Z),
 \label{eq:disc}\end{equation}
In literature a different value for the maximum value (Landau pole) is sometimes reported (about 4300\,GeV). We can trace the origin of this discrepancy to  the numerical value of $\alpha_{\rm em}^{-1}(M_Z)$ used in eq. \eqref{eq:disc}.

\begin{figure}[tbh!]
\centering
 \includegraphics[width=.9\linewidth]{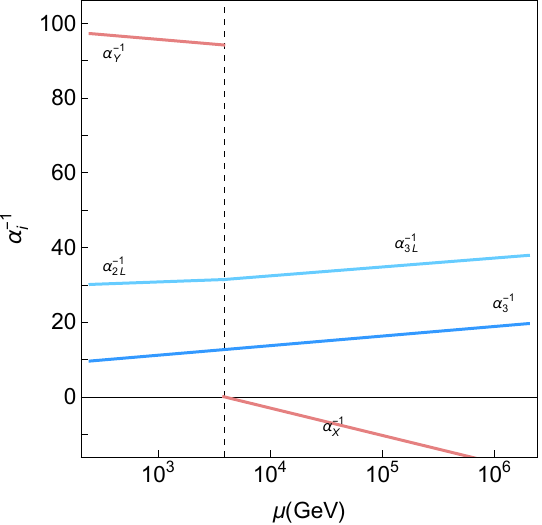}
\hfill
  \includegraphics[width=.9\linewidth]{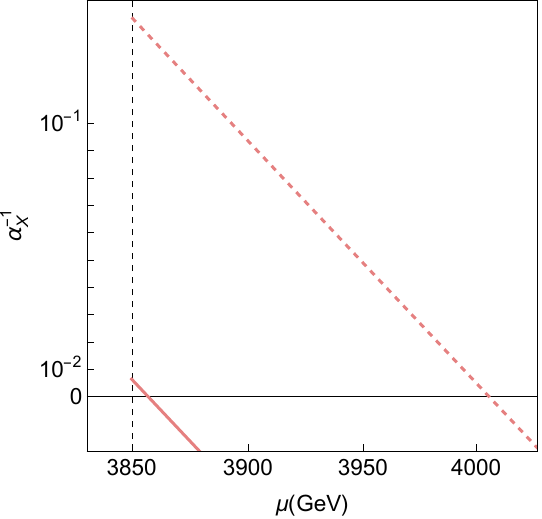}
  \caption{({\it Above}) Running of $\alpha_i^{-1}$ with the energy scale $\mu$.  The vertical dashed line is the arbitrary 331 breaking scale $\mu_{331}$. ({\it Below}) Zoom of $\alpha_X^{-1}$.  We compare with the multi-Higgs case with $N_\rho=4$  (dashed line) considered in section \ref{sec:Higgs}.}\label{fig2}
\end{figure}

\noindent
Let us now consider the range \eqref{e:range}    
and set the 331 breaking scale at the lower edge,  namely $\mu_{331}=3850$ \,GeV. 
 The running of $\alpha_X^{-1}(\mu)$ at energy scales $\mu>\mu_{331}$ is determined by its starting value $\alpha_X^{-1}(3850\,\mathrm{GeV})$, which in turn depend on $\beta$. This running also depends on
 the values of the $b_i$-parameters in eq. \eqref{e:coefficients}, which are 
\begin{equation}\label{b331above}
    b_X=22\,,\qquad
    b_{3L}= - 13/2\,,\qquad
    b_c= - 7\,.
\end{equation}\noindent
Below $\mu_{331}$,  the $b_i$-parameters are given by eq. \eqref{bSM2}.
In figure\,(\ref{fig2}) we plot the running of the coupling constants, above and below  $\mu_{331}$.
It is apparent  the dependence on  the matching conditions \eqref{eq:matching1}.
While  $\alpha_{2L}^{-1}$ run smoothly to $\alpha_{3L}^{-1}$  across $\mu_{331}$, there is  
a gap, which depends on $\beta$, between the values  $\alpha_Y^{-1}(\mu_{331})$ and $\alpha_X^{-1}(\mu_{331})$.
In the case $\beta=\pm\sqrt{3}$, we have 
 $\alpha_X^{-1}(3850\,\mathrm{GeV})=0.006$, which leads to 
 $\alpha_X^{-1}(\mu) = 0 $ (Landau pole) at $\mu \sim 3858$ GeV, not too far above $\mu_{331}$(see the right side of figure \,(\ref{fig2})).  This is not unexpected, given the bounds \eqref{e:range} on $\mu_{331}$ and the diverging behavior of the $U_X(1)$ coupling above that scale.
In the case
$\beta=\pm1/\sqrt{3}$ we would have had an higher starting point, $\alpha_X^{-1}(3850\,\mathrm{GeV})=13.951$, and a similar slope (because of  the same $b_i$), leading to an higher value for the Landau pole.\\ 
Let us say in passing that another common way to show the presence of the Landau pole for 331 extensions is through the weak mixing  angle $\theta_W$. It can be shown \cite{Buras:2012dp} that for $\beta = \sqrt{3}$ the following relation holds 
\begin{equation}
\frac{\alpha_X}{\alpha_{2L}} = \frac{6 \sin^2 \theta_W}{1 - 4 \sin^2 \theta_W}\,,\label{e:sinebsmbeta}
\end{equation}
which obviously links the divergence of $\alpha_X$ to $\sin^2\theta_W \to 1/4$, when $\alpha_{2L}$ is finite.
Thus one can use the running of $\sin^2 \theta_W$ to study the Landau   pole behavior
since
\begin{equation}
    \sin^2\theta_W = \frac{g_X^2}{6g^2_L + 4g^2_X} 
\end{equation}
The presence of a Landau pole at low energies poses a significant challenge, particularly in light of the stringent bounds on $\mu_{331}$ (hence on the bipleton's mass). The running in the $SU_L(3) \times U_X(1)$ sector is short-lived in terms of energy range.\\
The  mass values of bileptons, given by $m_Y \sim \mu_{331}/3$,  is close to the Landau pole. There is a narrowing gap between the areas where current experiments can look for  bileptons and  where the perturbation theory ceases to be accurate.

In order to rescue bileptons  in 331 models, it becomes necessary to move beyond the minimal 331 model. Several options can be considered, such as:
\begin{itemize}
    \item[1)] assuming light masses for the $D^1,D^2,T$ extra quarks;
    \item[2)] relaxing the condition $\xi^\ell_L=\ell_R^c$ and including exotic leptons;
    \item[3)] extending the scalar sector;  \item[4)] enlarging the number of families.
\end{itemize}
We have found that the cases 1) and 2) do not provide a viable solution--we find in both cases that the Landau pole is  again around 4 TeV. In the following sections we discuss the remaining two cases.

\section{Extending the scalar sector}
\label{sec:Higgs}
\begin{figure}[tbh!]
\centering
 \includegraphics[width=.9\linewidth]{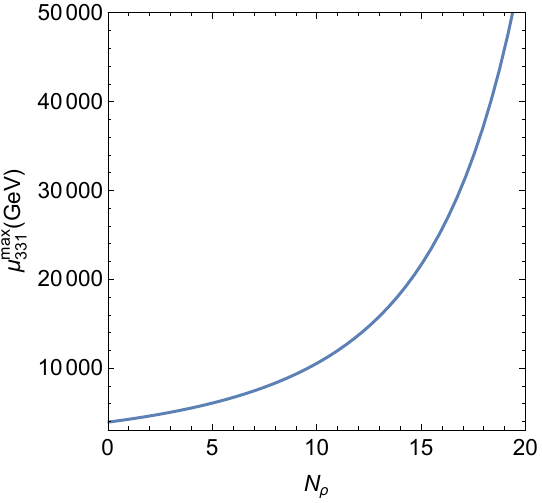}
  \caption{Maximal value of the 331 breaking scale as a function of the number of Higgs doublets.}\label{fignh}
\end{figure}

Let us consider extending
the scalar sector of the minimal 331 model 
through the inclusion of $SU_L(3)$ triplets, with the aim of shifting the maximal value for $\mu_{331}$. Since $m_Y \sim \mu_{331}$, the shift towards an higher value implies the possibility to have higher  bilepton masses in a regime which is still perturbatively valid.
\\ \noindent 
The light scalar triplet $\rho$ is neutral under $U_X(1)$ (see Table \ref{tab:minimal331}), hence the addition of an arbitrary number of $\rho$-like triplets  leaves $b_X$ invariant. 
This argument does not hold  for the other weak $b_i$-coefficients. 
In the minimal 331 model,  we add an arbitrary number of $\rho$-like scalar triplets $N_\rho$ which correspond to extra Higgs $SU_L(2)$ doublets and  contribute to the running between $\mu_{EW}$ and $\mu_{331}$. The $b_i$-parameters responsible of the running couplings below $\mu_{331}$ become (see  eqs. \eqref{formuleg} and  \eqref{bSM2}):
\begin{equation}\label{bSM3}
    b_Y=7+\frac{N_\rho}{6}\,,\qquad
    b_{2L}=-3+\frac{N_\rho}{6}\,,
    \end{equation}
At the maximum,  $N_\rho$ can be  $\tilde{N}_\rho =18$, since at that value $b_{2L}$ changes sign producing a new divergence in the coupling $g_L$ (at lowest order).
\noindent
 When  the $b_{2L}$ value moves from \eqref{bSM2}
to \eqref{bSM3}, 
$g_L(\mu)$ increases for every fixed  $\mu$ between the EW scale and $\mu_{331}$.
Above $\mu_{331}$,    eq.(\ref{b331above}) become:
\begin{equation}
    b_X=22\,,\qquad
    b_{3L}= - \frac{13}{2}+\frac{N_\rho}{6}\,,
\end{equation}
which leave $g_X(\mu)$ invariant.
The matching condition (\ref{eq:matching1}) gives the dependence of $\alpha_X^{-1}(\mu_{331}) $ on the scale $\mu_{331}$. Even if the running of $\alpha_X^{-1}$ above $\mu_{331}$ does not change, the 
different starting point 
 pushes the maximal value of $\mu_{331}$ ($\mu_{331}^{\mathrm{max}}$)  at higher energies. It can be seen explicitly in the right-hand side of figure (\ref{fig2}).
The dependence of $\mu_{331}^{\mathrm{max}}$ on $N_\rho$ is shown in figure\,(\ref{fignh}).

\noindent
 Summarizing,  the addition of $\rho$-like scalar triplets makes it possible to push $\mu^{\rm{max}}_{331}$ to higher energies, which allows  accommodating a wider range of mass values for the bileptons and  enlarging the energy range of validity (above $\mu_{331}$) of 331 extensions.
We highlight that the case with $N_H=2 + N_\rho=6$ can arise from an $E_6$ inspired framework. Indeed the 331 model can be embedded into $E_6$ as (see e.g. \,\cite{Slansky:1981yr})
\begin{align}
E_6 \to SU(2)\times &SU(6) \to  \nonumber \\
\to  &SU_c(3)\times SU_L(3) \times U(1)\,.
\end{align}
It is possible to show that the only colourless irreducible representations of $SU(6)$ are $\mathbf{6}$ and $\mathbf{84}$. These contain exactly six Higgs $SU_L(2)$ doublets. When $N_H=6$ we have $\mu_{331}^{\mathrm{max}}\sim 5500$ GeV.

\vskip5.mm
\noindent
There are other ways to extend the scalar sector.
So far, we have considered  one, two and many  light Higgs $SU_L(2)$ doublets that are embedded into $SU_L(3)$ triplets. The triplets  represent  minimal configurations for the scalar sector. Maintaining the natural assumption that 
fermions transform either as the representation ${\bf 3}$ or  ${\bf \bar 3}$  under $SU_L(3)$, gauge invariance also allows the possibility of  scalar sextets
(see e.g. \cite{Diaz:2003dk, Descotes-Genon:2017ptp, Addazi:2022frt}). It has to be noted that, should we choose to evade this assumption, e.g. including  octet leptons \cite{Dias:2004wk} or scalar leptoquarks assigned to the representation ${\bf 8}$  of $SU_L(3)$ \cite{Doff:2023bgy},  a wider choice of representations for the scalar sector would be available, and thus further ways to move the 
Landau pole. 

\begin{figure}[tbh!]
\centering
 \includegraphics[width=.9\linewidth]{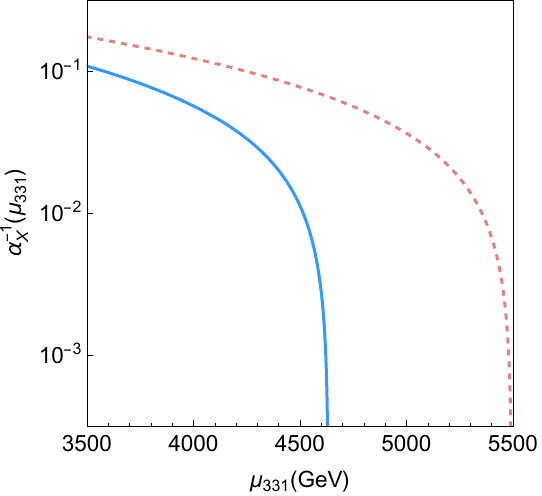}
  \caption{Running of $\alpha_X^{-1}$ in the sextet extension for different values of the mass of the doublet and triplet: electroweak scale ($dashed$ $line$), $600$ GeV (continuous line). }\label{figsextet}
\end{figure}

\noindent
Let us now extend  the scalar sector of the minimal 331 with a sextet $S$ of zero $X$-charge, which contributes to break spontaneously the 331 symmetry.
Its presence allows the addition of a renormalizable \footnote{Let us remark that there are  331 models where neutrino gets a Majorana mass by means of radiative contribution induced by triplet scalar representation, beside extra gauge bosons \cite{Singer:1980sw}.}  Majorana neutrino mass term\,\cite{Foot:1992rh}
\begin{equation}
    \mathcal{L}_Y = Y^{ij}\bar{L^c
}_i S L_j 
\end{equation}
where $L_i$ ($i= 1,2,3$) are the left-handed lepton $SU_L(3)$-triplets.
Since the sextet representation of $SU(3)$ decomposes into representations of the $SU(2)$ subgroup as $\mathbf{6} \rightarrow \mathbf{3} \oplus \mathbf{2} \oplus \mathbf{1}$, below $\mu_{331}$ we  have one triplet and  one iso-doublet and one scalar. 
In order to give a Majorana mass to neutrinos in the SM through the type II seesaw mechanism, one needs a $SU_L(2)$ scalar triplet to couple with lepton $SU_L(2)$ doublets. Following \cite{Doff:2023bgy} we assume 
$SU_L(2)$ doublet and triplet contained in the sextet $S$ to be light, while the singlet heavy.
The $b_i$-coefficients of eq.(\ref{formuleg}) are modified as
\begin{eqnarray}
    b_Y &=& \frac{20}{9} N_F + 
    \frac{1}{6} N_H + N_T\nonumber \\ 
    b_{2L} &=& -\frac{22}{3}+\frac{4}{3} N_F + 
    \frac{1}{6} N_H  +\frac{2}{3} N_T.\nonumber 
     \label{formuleg1}
\end{eqnarray}
We remind that $N_H$ is the number of (light) $SU_L(2)$ Higgs doublets and $N_T$ is the number of (light) $SU_L(2)$ Higgs triplets. In this case we have $N_H=3$ and $N_T=1$ and therefore
\begin{equation}
    b_Y=\frac{49}{6}\,,\qquad b_{2L}=-\frac{13}{6}\,,
\end{equation}
while $b_X$ is unchanged.
Using these $b_i$-parameters 
we obtain a value for the scale $\mu^{\rm max}_{331} \sim 5500$\,GeV as shown in  figure\,(\ref{figsextet}) with the dashed line.
However if we assume that the new fields belonging to the sextet participate to the running staring from an intermediate scale, let us say $\mu\sim 600$\,GeV, then 
$\mu^{\rm max}_{331} \sim 4630$\,GeV, see figure\,(\ref{figsextet}) continuous line.

\noindent

\section{4$th$ Family Extension Model}

As pointed out in section \ref{section31},  the Landau pole problem can be mitigated by enlarging the number of families.
 Here we extend the minimal model by adding  a 4$th$ generation in both the lepton and quark sector. 
It has been shown \cite{Eberhardt:2010bm,Eberhardt:2012gv,Djouadi:2012ae} that  the Standard Model with a sequential fourth generation (SM4), where the fourth family is an exact replica of the standard ones, is ruled out by data global fits. 
%This conclusion stems from the non-decoupling property of the SM4, where loop contributions involving fourth-generation fermions do not vanish as their masses increase.
%on processes involving heavy quarks in the gluon-gluon fusion loop. 
However these conclusions rely on the assumption of one single Higgs doublet and they do not apply to cases, like the present one, where we have  more (see e.g. \cite{Bellantoni:2012ag,Bar-Shalom:2012vvt,Das:2017mnu}).

In  331 models, anomaly cancellation requires that the matter assignment satisfies eqs.\,\eqref{eqal:sum} and\,\eqref{e:familynumber1}. Since eq.\,\eqref{e:familynumber} also holds,  the only possible configuration with $N_F=4$ is:
\begin{equation}
N^3_q = 2, N^{\bar{3}}_q = 2, N^3_l = 2, N^{\bar{3}}_l = 2.
\end{equation}
The matter content  is reported in Table \ref{tab:4fam}. 
\noindent
We observe that this 4$th$ family model includes exotic fermions beyond those present in the minimal   minimal 331 model (see Table\,\ref{tab:minimal331}), namely   quarks $b', t', T^2$ and  leptons $\tau',\nu_\tau', E^4$. Here we assume that $b', t', \tau', \nu_\tau'$ are light with masses below $\mu_{331}$, while 
$T^2, E^4$ are heavy 
with  masses which are proportional to the $\chi$ 
VEV. \\
\noindent
As far as we know, experimental constraints on the  sequential fourth family extending minimal 331 models are not available yet. Model-dependent bounds in different frameworks are given in \cite{Bar-Shalom:2011lgb,Mahapatra:2023zhi}.
%In order to provide experimental limits on exotic fermion $b',t',\tau'$ masses 
%a possibility is to 
Some insight can be gained by considering limits from vector-like quark searches. 
In particular we have
$m_{\tau'}>700$\,GeV\,\cite{CMS:2025urb}
 and 
$m_{b'}>1390\div1570$\,GeV\cite{CMS:2020ttz} 
from the CMS Collaboration (the range is due to  the variation of  branching ratios) and 
$m_{t'}>715\div950$\,GeV\,\cite{ATLAS:2015ktd} and $m_{b'}>1000\div2000$\,GeV\,\cite{ATLAS:2023ixh} from the ATLAS Collaboration.
A combined CMS and ATLAS study 
\cite{Benbrik:2024fku} reports  exclusion limits for  masses below  about 1500\,GeV of both $b'$ and $t'$.
We remark again that the previous limits cannot be  directly applied to our case.
In CMS and ATLAS analyses the lepton universality is  not required, while it is mandatory in a sequential fourth family model.\\
We note that requiring unitary at very high energies in $4th$ generation models,  suggests a bound on the BSM fermion masses  that is $m_f'\lesssim 500$\,GeV \cite{Chanowitz:1978uj}. However as reported in \cite{Djouadi:2012ae} this bound should not be viewed as a strict limit but simply as an indication that strong dynamics takes place. Moreover simulation of a strong Yukawa coupling regime on the lattice found $m_f'\lesssim 700$\,GeV \cite{Gerhold:2010wv}. Therefore hereafter we will consider the special scale $m_{\rm NP}\sim 700$\,GeV as a benchmark case.

\begin{figure}[tbh!]
\centering
 \includegraphics[width=.9\linewidth]{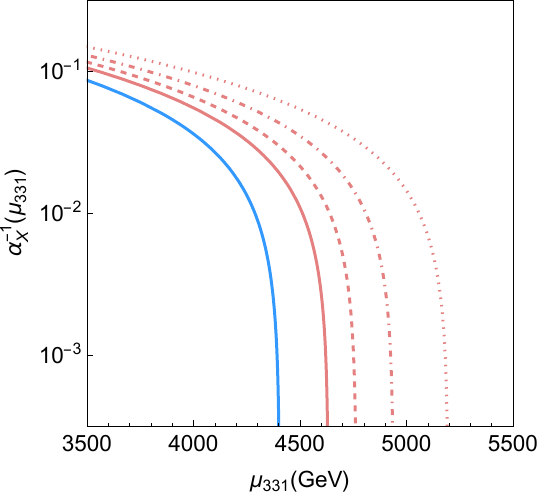}\\
\includegraphics[width=.9\linewidth]{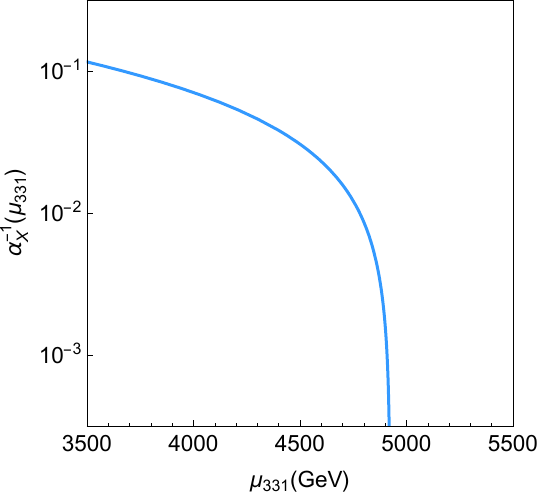}
  \caption{Running of $\alpha_X^{-1}$ with the energy scale $\mu_{331}$. ({\it Top plot}) 4th family model for  different values of the mass scale $m_{\rm NP}$: 1500 GeV (dotted), 1000 GeV (dot-dashed), 700 GeV (dashed), 500 GeV (continue). ({\it Bottom plot}) 4th family model + sextet with $m_{\rm NP} \sim 700$\,GeV.}\label{fig4family2}
\end{figure}

\noindent
For the sake of simplicity, we assume $b', t', T^2$ to have the same mass $m_{\rm NP}$ and we consider e.g. the cases where $m_{\rm NP}$=500, 700, 1000, 1500 GeV.
%In order to find $\mu^{\rm max}_{331}$ 
We  proceed in two steps: we run $\alpha^{-1}_X(\mu)$ from the electroweak breaking scale to $m_{\rm NP}$ using the $b_i$-parameters\,(\ref{bSM2}). Then we run it above $m_{\rm NP}$ using the following $b_i$-parameters:
%(\ref{e:coefficients})
\begin{eqnarray}
b_Y &=& \frac{83}{9}, \quad b_{2L} = - \frac{5}{3} \qquad \text{for $\mu <\mu_{331}$},\label{b4f1}\\
b_X &=& \frac{134}{3}, \quad b_{3L} = - \frac{31}{6} \quad \text{for $\mu >\mu_{331}$}. \label{b4f2}
\end{eqnarray}
By using these parameters 
we determine the dependence of  $\alpha_X^{-1}$ on $\mu_{331}$ for different values of the mass scale $m_{\rm NP}$. This is shown in the upper plot of figure\,(\ref{fig4family2}). 
\noindent
In the benchmark case $m_\text{NP} \sim 700$\,GeV,  we find $\mu^{\text{max}}_{331} \sim 4840$ GeV.\\
As a final step we also include the scalar sextet (see section \ref{sec:Higgs}), obtaining $\mu^{\text{max}}_{331} =5912$\,GeV as shown in the lower plot of figure\,(\ref{fig4family2}).\\

\vskip5.mm
In Table \ref{btab}, we compare the different extensions of the minimal 331 models we examined. We report the values of $b_i$-parameters in the energy range between the electroweak and the 331 symmetry breaking scales, the values of $\mu^{\rm{max}}_{331}$ and the  bileptons' masses estimated at $\mu^{\rm{max}}_{331}$, according to eq. \eqref{e:bileptmass}.

\begin{table}[h!]
    \centering
    \begin{tabular}{|c|c|c|c|c|}
    \hline
           & $b_Y$ & $- b_{2L}$ & $\mu_{\rm max}$(GeV) & $m_Y$(GeV) \\
           \hline
        minimal & 7 & 3 & 3908 & 1270\\
        sextet & 49/6& 13/6  & 4531 & 1472\\
 $N_H=6$ & 47/6 & 13/6 & 5500 & 1788\\
        4$th$ f. & 83/9 & 5/3 & 4840 & 1573 \\
        4$th$ f.+sext & 187/18 & 5/6 & 5912&1921 \\
        \hline
    \end{tabular}
    \caption{Summary of $b$-parameters and corresponding values of the Landau pole for different 331 models and expected bilepton mass:
    minimal 331, only sextet, extra $\rho$-like triplets  ($N_H=6$),
    4$th$ family and 4$th$ family + sextet with $\text{m}_\text{NP} \sim 700$ GeV.     }
    \label{btab}
\end{table}

\begin{figure}[tbh!]
\centering
 \includegraphics[width=\linewidth]{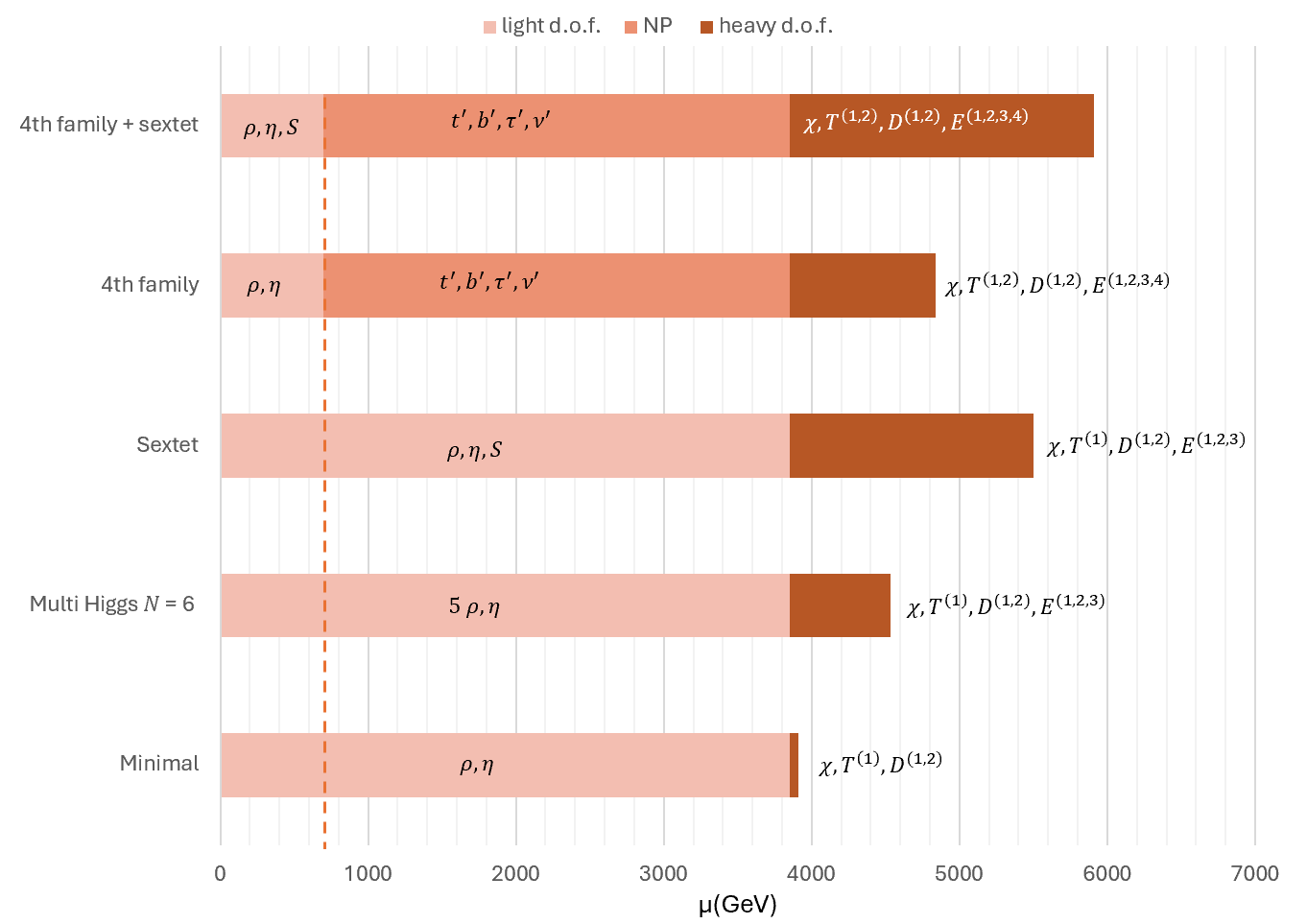}\\
  \caption{Benchmark models running report}\label{runningreport}
\end{figure}

\noindent

\section{Conclusions}
The 331 extensions of the Standard Model are characterized by a free parameter $\beta$. By setting $\beta=\sqrt{3}$, 331 models predict doubly charged gauge bosons that mainly decay into pairs of same-sign leptons and are therefore dubbed bileptons. Their production and decay yield distinctive signatures that enable direct searches at colliders.\\
\noindent
The minimal realizations of 331 extensions with $\beta=\sqrt{3}$ are plagued by the presence of a Landau pole around 4\,TeV, which significantly constrains their range of perturbative validity.  
The bilepton mass is predicted to be very close to 1300\,GeV, a value that will soon be either observed or ruled out by the LHC. 
Assuming the worst-case scenario of no detection, we investigate whether bileptons remain viable within the 331 framework. To answer this question, we address the issue of the Landau pole at approximately 4\,TeV and explore possible ways to shift the pole to higher energy scales. This would enlarge the range over which the spontaneous symmetry breaking of the 331 model can occur and thus rescue bileptons, allowing for higher mass values.\\ 
\noindent
We begin by analyzing the running of the gauge coupling constants in the minimal 331 model, considering its matter content and the associated scales of spontaneous symmetry breaking for both the 331 and electroweak sectors.
We find that, to comply with experimental constraints, this model must behave as a Two Higgs Doublet Model in the energy range between $\mu_{\rm{EW}}$ and $\mu_{331}$, with both Higgs bosons acquiring mass within this range. Unfortunately, we also find that it is
not possible to shift the Landau pole by varying the free parameters, namely $\mu_{331}$ and the masses of the exotic quarks. This remains true even in the so-called next-to-minimal 331 model, which includes additional non-standard leptons.\\
Viable solutions require
extending the scalar sector and/or adding a sequential fourth family of non-standard quarks and leptons. 
The running below $\mu_{331}$ of the $U_Y(1)$ and $SU(2)_L$ coupling constants changes upon supplementing the minimal model with $\rho$-like Higgs triplets (corresponding to extra Higgs $SU_L(2)$ doublets). 
Although this procedure is constrained by bounds on the behavior of the $SU_L(2)$ coupling constant, the addition of scalar triplets of this kind extends the valid energy range, which in turn allows for a wider spectrum of bilepton masses, e.g. $m_Y \sim 1788$ GeV if $N_H=6$. Notably, the case with $N_H=6$
can originate from an $E_6$-inspired framework.\\
Another scalar sector extension analyzed in this work includes an $SU_L(3)$ sextet. This case, which allows the introduction of a mass term for Majorana neutrinos, predicts $m_Y \sim 1472 $  GeV.\\
Finally, we have studied an extension of the fermionic sector involving a sequential fourth family with two Higgs doublets. 
%This extension is not forbidden in the context of Two Higgs Doublet Models. 
By making conservative assumptions on the mass of the fourth family based on recent experimental results, one manages to shift the Landau pole, and get $m_Y \sim 1573 $\,GeV with an intermediate mass scale of 700 GeV.  For the sake of completeness,  we also study this model including  the  scalar sextet, obtaining the higher value $m_Y \sim 1921 $\,GeV at the same intermediate scale.\\
\noindent 
We summarize our results 
%of  the benchmark models
 in Table \ref{btab} and Figure \ref{runningreport}. Extending the minimal 331 models  increases the Landau pole, yet this increase remains within the range of experimental values attainable at LHC in the near future.

\vspace*{-0.6cm}
\section*{Acknowledgements} 
The authors thank G. Corcella and P.H. Frampton for interesting discussions. The work of G.R. was partially supported by the research project ENP (Exploring New physics) funded by INFN. S.M acknowledges the partial support by the research project TAsP (Theoretical Astroparticle Physics) funded by INFN. G.P.P. acknowledges the partial support by the research project TCCP (Theoretical Particle Physics and Cosmology) funded by INFN.
G.R. and G.P.P. thank the Galileo Galilei Institute for Theoretical Physics for hospitality during the early stage of this work.\\

\newpage
\appendix
\section*{Tables of matter content of the models}

\begin{table}[tbh!]
    \centering
%    \begin{minipage}{0.35\textwidth}
%    \centering
%    \renewcommand{\arraystretch}{1.6} % Adjusts the row height
    \begin{tabular}{|c|c|c|c|c|}
        \hline
        && $Q$ & $SU_{L}(3)$ & $U_X(1)$\\
        \hline
        \multirow{3}{*}{Quarks L}
        & $u_L$ & 2/3 &  &  \\
        & $d_L$ & -1/3 & 3 & -1/3 \\
        & $D^1_L$ & -4/3 &  &  \\
          \cline{2-5}
        & $c_L$ & 2/3 &  &  \\
        & $s_L$ & -1/3 & 3 & -1/3 \\
        & $D^2_L$ & -4/3 &  &  \\
          \cline{2-5}
        & $b_L$ & -1/3 &  &  \\
        & $-t_L$ & 2/3 & $\bar{3}$ &  2/3 \\
        & $T_L$ & 5/3 &  &  \\
        \hline
\multirow{3}{*}{Leptons L}
        & $\ell^{}_L$ & $-1$ &  &  \\
        & $-\nu^{\ell}_L$ & 0 & $\bar{3}$ &  0 \\
        & $\ell_R^c$ & $+1$ &  &  \\
        \hline
%    \end{tabular}
%       %  \caption*{Table 1: Description for Table 1}
%    \end{minipage}
%    \hfill
%    \begin{minipage}{0.50\textwidth}
%        %\centering
%         \hspace*{1cm}
%         \begin{tabular}{|c|c|c|c|c|}
%        \hline
%         && $Q$ & $SU_{L}(3)$ & $U_X(1)$\\
%        \hline
        \multirow{3}{*}{Quarks R} 
        & $u_R,c_R$ & 2/3 &  & 2/3  \\
        & $d_R,s_R$ & $-1/3$ &  &  $-1/3$ \\
        & $D^1_R,D^2_R$ & $-4/3$ & 1 & $-4/3$  \\
        & $b_R$ & $-1/3$ &  & $-1/3$  \\
        & $t_R$ & 2/3 & $ $ &  2/3 \\
        & $T_R$ & 5/3 &  & 5/3  \\
        \hline
    \end{tabular}
%    \newline
%\vspace*{1cm}
%\newline
%   \hspace*{1cm}
%         \begin{tabular}{|c|c|c|}
%        \hline
%        Scalar & $SU_L(3)$ & $U_X(1)$ \\
%        \hline 
%        $\chi$ & 3 &  $+1$\\
%        $\rho$ & 3 & 0\\
%        $\eta$ & 3 & $-1$\\
%        \hline
%    \end{tabular}
%    \end{minipage}
    \caption{Matter content of the minimal 331 model with $\beta = \sqrt{3}$. }
    \label{tab:minimal331}
\end{table}

\begin{table}[tbh!]
    \centering
%    \begin{minipage}{0.35\textwidth}
%    \centering
%    \renewcommand{\arraystretch}{1.6} 
    \begin{tabular}{|c|c|c|c|c|c|}
        \hline
        && $Q$ & $SU_{L}(3)$ & $U_X(1)$& $U_Y(1)$\\
        \hline
        \multirow{3}{*}{Quarks L} 
        & $u_L$ & 2/3 &  &  & 1/6\\
        & $d_L$ & -1/3 & 3 & -1/3 &1/6\\
        & $D^1_L$ & -4/3 &  & & -4/3\\
          \cline{2-6}
        & $c_L$ & 2/3 &  & & 1/6\\
        & $s_L$ & -1/3 & 3 & -1/3 & 1/6\\
        & $D^2_L$ & -4/3 &  & & -4/3\\
          \cline{2-6}
        & $b_L$ & -1/3 &  &  &1/6\\
        & $-t_L$ & 2/3 & $\bar{3}$ &  2/3& 1/6\\
        & $T^1_L$ & 5/3 &  &  &5/3\\
            \cline{2-6}
        & $b'_L$ & -1/3 &  &  &1/6\\
        & $-t'_L$ & 2/3 & $\bar{3}$ &  2/3& 1/6\\
        & $T^2_L$ & 5/3 &  &  &5/3\\
        \hline
\multirow{3}{*}{Leptons L}
 & $\nu_{eL}$ & 0 &  &  &-1/2\\
        & $e_L$ & -1 & $3$ &  -1 &-1/2\\
        & $E^1_L$ & -2 &  &  &-2\\
         \cline{2-6}
          & $\nu_{\mu L}$ & 0 &  & & -1/2\\
        & $\mu_L$ & -1 & $3$ &  -1 &-1/2\\
        & $E^2_L$ & -2 &  &  &-2\\
         \cline{2-6}
        & $\tau_L$ & $-1$ &  &  &-1/2\\
        & $-\nu_{\tau L}$ & 0 & $\bar{3}$ &  0 &-1/2\\
        & $E^3_L$ & $+1$ &  &  &+1\\
                 \cline{2-6}
        & $\tau'_{L}$ & $-1$ &  &  &-1/2\\
        & $-\nu'_{\tau L}$ & 0 & $\bar{3}$ &  0 &-1/2\\
        & $E^4_L$ & $+1$ &  &  &+1\\
        \hline
%    \end{tabular}
%      
%    \end{minipage}\hfill
%    \begin{minipage}{0.50\textwidth}
%        %\centering
%         \hspace*{1cm}
%         \begin{tabular}{|c|c|c|c|c|}
%        \hline
%         && $Q$ & $SU_{L}(3)$ & $U_X(1)$\\
%        \hline
        \multirow{3}{*}{Quarks R} 
        & $u_R,c_R$ & 2/3 &  & 2/3 & \\
        & $d_R,s_R$ & $-1/3$ &  &  $-1/3$& \\
        & $D^1_R,D^2_R$ & $-4/3$ & 1 & $-4/3$ & \\
        & $b_R,b'_R$ & $-1/3$ &  & $-1/3$ & \\
        & $t_R,t'_R$ & 2/3 & $ $ &  2/3 &\\
        & $T^1_R,T^2_R$ & 5/3 &  & 5/3 & \\
        \hline
        \multirow{3}{*}{Leptons R}
        & $\nu_{eR},\nu_{\mu R}$ & 0 &  & 0&\\
        & $e_R,\mu_R$ & -1 &  & -1 & \\
        & $E^1_R,E^2_R$ & -2 & 1 & -2&\\
        & $\nu_{\tau R},\nu'_{\tau R}$ & 0 &  & 0&\\
        & $\tau_R,\tau'_{R}$ & -1 &  & -1 & \\
        & $E_R^3,E_R^4$ & 1 &  & 1&\\
        \hline
    \end{tabular}
%    \newline
%\vspace*{1cm}
%\newline
%   \hspace*{1cm}
%         \begin{tabular}{|c|c|c|}
%        \hline
%        Scalar & $SU_L(3)$ & $U_X(1)$ \\
%        \hline 
%        $\chi$ & 3 &  $+1$\\
%        $\rho$ & 3 & 0\\
%        $\eta$ & 3 & $-1$\\
%        \hline
%    \end{tabular}
%    \end{minipage}
    \caption{Matter content of the 331 model with 4th family.
%    : we use the notation of Ref.\,\cite{Eberhardt:2012gv}. 
    }
    \label{tab:4fam}
\end{table}
\newpage

\nocite{*}
\bibliographystyle{unsrt} 
\bibliography{draftPRD}

\newpage
%\begin{figure}[tbh!]
%\centering

%\includegraphics[width=2.1\linewidth]{331_regime.pdf}
%  \caption{Ranges of the scale  $\mu_{331}$ where there are indicated the fields participating in the running between the electroweak and the 331 scale. All other particles which are not indicated are considered to have mass at (or above) the $\mu_{331}$ scale.}\label{regime}
%\end{figure}
\end{document}